\documentclass[11pt]{article}

\usepackage{amsmath}
\usepackage{amsfonts}
\usepackage{amssymb}
\usepackage{amscd}
\usepackage{latexsym}
\usepackage{mathrsfs}
\usepackage{amsthm}

\newtheorem{Theorem}{Theorem}
\newtheorem{Remark}{Remark}

\def\C{\mathbb{C}}
\def\de{\mathrm d}

\def\R{\mathbb{R}}
\def\S{\mathbb{S}}

\begin{document}

\title{Levinson's theorem for Schr\"odinger operators with 
point interaction: a topological approach} 

\author{Johannes Kellendonk\,~and\,~Serge Richard}
  \date{\small
    \begin{quote}
      \emph{
    \begin{itemize}
    \item[]
                        Institut Camille Jordan,
                        B\^atiment Braconnier,
                        Universit\'e Claude Bernard Lyon 1, \\
                        43 avenue du 11 novembre 1918,
                        69622 Villeurbanne cedex, France
    \item[]
      \emph{E-mails\:\!:}
      kellendonk@math.univ-lyon1.fr\,~and\,~srichard@math.univ-lyon1.fr
    \end{itemize}
      }
    \end{quote}
    July 2006
  }

\maketitle

\begin{abstract}
In this note Levinson theorems for Schr\"odinger operators in $\R^n$
with one point interaction at $0$ are derived using the concept of
winding numbers. These results are based on new expressions for the associated
wave operators.
\end{abstract}

\section{Introduction}
In \cite{KR} we proposed to look afresh at Levinson's theorem by
viewing it as an identity between topological invariants, one
associated with the bound state system, the other associated with the
scattering system. 
Here we present the complete analysis for a class of solvable 
models in quantum mechanics
which goes under the name of one point interaction at the origin 
($\delta$ and $\delta'$ interactions). 
For these models we find novel expressions for the wave operators 
which allow us to prove our topological version of Levinson's theorem
and to exhibit our ideas without
recourse to techniques from algebraic topology.

What we prove is the following:
Let $H$ be a Schr\"odinger operator describing a $\delta$ interaction at $0$ in $\R^n$
for $n \in \{1,2,3\}$ or a $\delta'$ interaction at $0$ in $\R^1$ as
discussed very carefully in \cite[Chapter I]{AGHH}.
The wave operator $\Omega_-$ for the couple $(H, -\Delta)$ 
can be rewritten in the form
\begin{equation*}
\Omega_- - 1 = \varphi(A)\eta(-\Delta) P 
\end{equation*}
where  $A$ is the generator of dilation in $\R^n$, $P$ is an
appropriate projection, and $\varphi,\eta$ are continuous functions which have
limits at the infinity points of the spectra $\sigma(A)$ and
$\sigma(-\Delta)$, respectively, i.e.\ 
$\displaystyle{\lim_{t \to \pm\infty}\varphi(t)}$ and
$\displaystyle{\lim_{t\to 0, +\infty}\eta(t)}$ exist. This allows us to
define a continuous function $\Gamma:B\to \C^*$, from the boundary $B$ of the
square $(\sigma(-\Delta)\cup\{0,+\infty\}) \times (\sigma(A)\cup\{-\infty,+\infty\})$ 
by setting
\begin{equation*}
\Gamma(\epsilon,a) = \varphi(a)\eta(\epsilon)+1,\quad (\epsilon,a)\in B.
\end{equation*}
Since $B$ is topologically isomorphic to a circle, $\Gamma$ has a winding number
$w(\Gamma)$ which is defined as the number of times $\Gamma(t)$
wraps (left) around $0\in\C$ when $t$ goes around $B$. 
This requires a choice of orientation for $B$ which we fix as follows:
$B$ consists of four sides of the square one of which is
$B_2=(\sigma(-\Delta)\cup\{0,+\infty\})\times \{+\infty\}\cong[0,+\infty]$; we
choose on $B$ the orientation which corresponds on $B_2$ to the natural order
on $[0,+\infty]$.
Our main result states a relation between
this number and the number of bound states of $H$ which is  
 $\# \sigma_{\hbox{\rm \tiny p}}(H)$.
\begin{Theorem}[Levinson's theorem for $\delta$ and $\delta'$ interaction]
Let $H$ be a Schr\"odinger operator defined by a $\delta$-interaction at $0$ in
$\R^n$ with $n \in \{1,2,3\}$ or by a $\delta'$-interaction at $0$ in $\R^1$. 
Then
\begin{equation*}
w(\Gamma) = - \# \sigma_{\hbox{\rm \tiny p}}(H).
\end{equation*}
\end{Theorem}
We prove this result in the next section by explicit verification.
It is in fact a special case of an index theorem \cite{CH,KRinfty}.

Let us provide a few words of explanation for $w(\Gamma)$. Assuming
differentiability it can be calculated by the integral
\begin{equation*}
w(\Gamma) = \frac{1}{2\pi i}\int_B \Gamma^{-1} \de \Gamma\ .
\end{equation*}
To interpret this expression it is
convenient to consider the four sides
$B_1=\{0\}\times(\sigma(A)\cup\{-\infty,+\infty\})$,
$B_2= (\sigma(-\Delta)\cup\{0,+\infty\})\times\{+\infty\}$,
$B_3= \{+\infty\} \times (\sigma(A)\cup\{-\infty,+\infty\})$,
$B_4=(\sigma(-\Delta)\cup\{0,+\infty\})\times \{-\infty\} $
of the square. Then
\begin{equation*}
w(\Gamma) = \sum_{i=1}^4 w_i,\quad 
w_i= \frac{1}{2\pi i} \int_{B_i} \Gamma^{-1}_i
\de \Gamma_i
\end{equation*}
where $\Gamma_i$ is the restriction of $\Gamma$ to $B_i$. 
It can be observed for all the following examples that 
$\Gamma_2(-\Delta)P +P^\perp $ is equal to the scattering operator
$S$ and that $\Gamma_4(-\Delta) = 1$. This behaviour is not a coincidence but must 
hold in general \cite{KRinfty}. In other words $w(\Gamma)$ contains as contribution the term
\begin{equation*}
w_2 =
\frac{1}{2\pi i} \int_0^{+\infty} \mbox{\rm tr} [S^*(\epsilon)S'(\epsilon)] \de \epsilon \ ,
\end{equation*} 
where ${\rm tr}$ is the usual trace on the compact operators of $L^2(\S^{n-1})$.
Now $w_2$ is the integral over the time delay one usually finds in
Levinson's theorem. In dimension $n = 2$ one even has
$\Gamma_i(-\Delta) = 1$ for all $i\neq 2$ and so there
is no other contribution to $w(\Gamma)$.
But for most of the other examples, the low and the high energy
behaviour of the wave operator is non-trivial leading to
contributions of $\Gamma_1$ and $\Gamma_3$ to the winding number.

Expressions relating $w_2$ to the number of bound states for one point
interactions can be found in the physics literature, see e.g.\
\cite{Graham}, usually providing different arguments for the
occurences of corrections. Here these corrections appear as the
missing parts ($w_1$ and $w_3$) of a winding number calculation. This
not only gives a full and coherent explanation of these corrections
but also makes clear that Levinson's theorem is of topological nature. 
In a future publication \cite{KRinfty}, we shall show that a similar 
picture holds for general Schr\"odinger operators and that the
corrections added in such  context can also be fully explained.

The proof the theorem as well as more explanations on the underlying 
constructions are given in the following sections. 
New formulas for the wave operators
are introduced successively for the $\delta$-interaction in $\R^n$ for
$n=3$, $2$ and $1$,  
and then for the $\delta'$-interaction in $\R^1$. 

\section{Schr\"odinger operators with point interaction}

A so-called Schr\"odinger operator with a $\delta$ or a $\delta'$
interaction at the origin can be defined as a
self-adjoint extension of the restriction of the Laplace operator 
to a suitable subspace of $L^2(\R^n)$. These operators are discussed
in great detail in  \cite[Chapter I]{AGHH}. 
 
Common to all the self-adjoint extensions $H$ 
we look at here is that
\begin{equation*}
\sigma_{\hbox{\rm \tiny ess}}(H) = \sigma_{\hbox{\rm \tiny ac}}(H) 
= [0,\infty)\ , \quad \sigma_{\hbox{\rm \tiny sc}}(H)=\emptyset\ .
\end{equation*}
The point spectrum of $H$, however, depends on both the extension and the dimension.
The main feature of these models is that the wave operators defined by
\begin{equation*}
\Omega_\pm :=
s-\lim_{t\to \pm \infty}e^{itH}\;\!e^{-it(-\Delta)}
\end{equation*}
can be explicitly computed. We recall their explicit form, which
depends on the extension and the dimension, further down.
A common property shared 
by these wave operators is that they act non-trivially only on a small 
subspace of $L^2(\R^n)$. Denoting by $P$ the orthogonal 
projection onto that subspace we have the possibility that $P=P_0$ 
the orthogonal projection onto the rotation invariant subspace of $L^2(\R^n)$,
or, for $n=1$, $P=P_1$, the orthogonal
projection $P_1$ onto the antisymmetric functions of $L^2(\R)$. 
Recall that the ranges of $P_0$ or $P_1$ are invariant under the usual 
Fourier transform in $L^2(\R^n)$. 

The dilation group in $\R^n$ and its generator $A$ play an important
role. We recall that
its action on $\psi \in L^2(\R^n)$ is given by $\big(U(\theta)\psi\big)(x)=
e^{n\theta/2}\psi(e^\theta x)$ for all $x \in \R^n$ and $\theta \in \R$, and 
that its self-adjoint generator $A$ has the 
form $\hbox{$\frac{1}{2i}$}(Q\cdot \nabla + \nabla \cdot Q)$ on
$C_0^\infty(\R^n)$. 
The group leaves the range of $P_0$ and $P_1$ invariant. 
We refer to \cite{Jensen} for more information about this group 
and for a detailed description of the Mellin transform 
$\mathcal M$ which is a unitary transformation diagonalizing $A$.

\subsection{The dimension $n=3$}

The operator $-\Delta$ defined on $C_0^\infty(\R^3 \setminus \{0\})$ 
has deficiency indices $(1,1)$ so that 
all its self-adjoint extensions $H_\alpha$ can be parametrized 
by an index $\alpha$ belonging to $\R\cup\{\infty\}$. This parameter
determines a certain boundary condition at $0$ but $-4\pi\alpha$
also has a physical
interpretation as the inverse of the scattering length.
The choice $\alpha=\infty$ corresponds to the free Laplacian
$-\Delta$. 
$H_\alpha$ has a single bound state for $\alpha <0$ 
at energy $-(4\pi\alpha)^2$ but
no point spectrum for $\alpha \in [0,\infty]$. 
Furthermore, the action of the wave 
operator $\Omega_-^\alpha$ for the couple $(H_\alpha, -\Delta)$ 
on any $\psi \in L^2(\R^3)$ is given by:
\begin{equation*}
\big[(\Omega_-^\alpha-1)\psi\big](x)=s-\lim_{R\to \infty}(2\pi)^{-3/2}
\int_{k\leq R}k^2 \;\!\de k \int_{\S^2} \de \omega \;\!\frac{e^{ik|x|}}{(4\pi\alpha -ik)|x|}
\;\!\hat{\psi}(k\omega)\ ,
\end{equation*}
where $\hat{\psi}$ is the $3$-dimensional Fourier transform of $\psi$. 

Now, one first easily observes that 
$\Omega_-^\alpha-1$ acts trivially on the orthocomplement of the range
of $P_0$. One may also notice that it can be rewritten as a product of three operators, {\it i.e.}
$\Omega_-^\alpha -1 = T_2 T_1 P_0$ with 
\begin{equation*}
T_1 = \frac{2i\sqrt{-\Delta}}{4\pi\alpha -i\sqrt{-\Delta}}
\end{equation*}
and
\begin{equation*}
\big[T_2\psi\big](x)=s-\lim_{R\to \infty} (2\pi)^{-1/2}
\int_{k\leq R}k^2 \;\!\de k \;\!\frac{e^{ik|x|}}{i k|x|}
\;\!\hat{\psi}(k)\ . 
\end{equation*} 
Finally, let us observe that $T_1+1$ is simply equal to the scattering operator 
$S^\alpha:=(\Omega_+^\alpha)^* \;\!\Omega_-^\alpha$ and that $T_2$ is invariant 
under the action of the dilation group: $U(\theta)T_2U(-\theta) = T_2$.
Thus, $T_2$ can be diagonalized in the spectral representation of $A$. 
A direct calculation using the expression for $\mathcal M$ from 
\cite{Jensen} leads to the following result:
\begin{equation*}
\Omega_-^\alpha-1 = \left[\frac{1}{2}\left(1+\tanh(\pi A)-i\big(\cosh(\pi A)\big)^{-1}\right)\right]
\left\{\frac{2i\sqrt{-\Delta}}{4\pi\alpha -i\sqrt{-\Delta}}\right\} P_0\ .
\end{equation*}
So let us set
\begin{equation*}
 r(\xi) =  -\tanh(\pi \xi) +i\big(\cosh(\pi \xi)\big)^{-1}\; .
\end{equation*}
and 
\begin{equation*}
s^\alpha(\xi) =  \frac{4\pi\alpha +i\sqrt{\xi}}{4\pi\alpha -i\sqrt{\xi}}\ 
\end{equation*}
As a consequence of the expression for $\Omega_-^\alpha -1$ the functions $\Gamma_i$ and
their contributions to the winding number are given by 
\begin{center}
\begin{tabular}{|c|c|c|c|c|c|c|c|c|c|}
\hline
 & $\Gamma_1$ & $\Gamma_2$ & $\Gamma_3$ & $\Gamma_4$  & $w_1$
& $w_2$ & $w_3$ & $w_4$ & $w(\Gamma)$ \\ \hline\hline
$\alpha < 0$ &$ 1 $&$ s^\alpha $&$ r $&$ 1 $&$ 0 $&$ -\frac{1}{2}  $&$
-\frac{1}{2}$&$0 $&$ -1 $ \\\hline
$\alpha = 0  $&$  r $&$ -1 $&$ r $&$ 1 $&$ \frac{1}{2}  $&$0$& $
-\frac{1}{2}$&$0 $&$ 0 $ \\\hline
$\alpha > 0  $&$ 1 $&$ s^\alpha $&$ r $&$ 1 $&$ 0 $&$ \frac{1}{2}  $&$
-\frac{1}{2}$&$0 $&$ 0 $ \\
\hline 
$\alpha = \infty  $&$ 1 $&$ 1 $&$ 1 $&$ 1 $&$ 0 $&$ 0  $&$0$&$0 $&$ 0 $ \\
\hline
\end{tabular}
\end{center}
and we see that  $w(\Gamma)$ corresponds to minus the number of bound states of $H_\alpha$.

\subsection{The dimension $n=2$}

The situation for $n=2$ parallels that of $n=3$ in that
the operator $-\Delta$ defined on $C_0^\infty(\R^2 \setminus \{0\})$ 
has deficiency indices $(1,1)$ and that 
all its self-adjoint extensions $H_\alpha$ can be parametrized 
by an index $\alpha$ belonging to $\R\cup\{\infty\}$. Again $\alpha$ 
determines a certain boundary condition at $0$ and $-2\pi\alpha$
has a physical interpretation as the inverse of the scattering length.
Also here the choice $\alpha=\infty$ corresponds to the free Laplacian
$-\Delta$. But in two dimensions  $H_\alpha$ has a single eigenvalue 
for all $\alpha\in\R$. The wave operator
$\Omega_-^\alpha$ for the couple $(H_\alpha, -\Delta)$ acts on $\psi \in
L^2(\R^2)$ as     
\begin{equation*}
\big[(\Omega_-^\alpha-1)\psi\big](x)=s-\lim_{R\to \infty}(2\pi)^{-1}
\int_{k\leq R}k \;\!\de k \int_{\S^1} \de \omega \;\!\frac{i\pi/2}{2\pi\alpha-\Psi(1)+\ln(k/2i)}H_0^{(1)}(k|x|)
\;\!\hat{\psi}(k\omega)\ ,
\end{equation*}
where $\hat{\psi}$ is the $2$-dimensional Fourier transform of $\psi$,
$H_0^{(1)}$ denotes the Hankel 
function of the first kind and order zero, and $\Psi$ is the digamma function.
A similar calculation as above yields that this wave operator 
can be rewritten as
\begin{equation*}
\Omega_-^\alpha-1 = \left[\frac{1}{2}\big(1+\tanh(\pi A/2)\big)\right]
\left\{\frac{i \pi}{2\pi\alpha -\Psi(1) +
    \ln(\sqrt{-\Delta}/2)-i\frac{\pi}{2}}\right\} P_0\ . 
\end{equation*}
Thus we get the following results for the functions $\Gamma_i$ and
their contribution to the winding number. Let us set
\begin{equation*}
r(\xi)=-\tanh(\pi \xi / 2)
\end{equation*}
and
\begin{equation*}
s^\alpha(\xi) =  
\frac{2\pi\alpha -\Psi(1) + \ln(\sqrt{\xi}/2)+i\pi/2}
{2\pi\alpha -\Psi(1) + \ln(\sqrt{\xi}/2)-i\pi/2}\ .
\end{equation*}
Then
\begin{center}
\begin{tabular}{|c|c|c|c|c|c|c|c|c|c|}
\hline
 & $\Gamma_1$ & $\Gamma_2$ & $\Gamma_3$ & $\Gamma_4$  & $w_1$
& $w_2$ & $w_3$ & $w_4$ & $w(\Gamma)$ \\ \hline\hline 
$\alpha \in\R $ &$ 1 $&$ s^\alpha $&$ 1 $&$ 1 $&$ 0 $&$ -1  $&$
0 $&$0 $&$ -1 $ \\\hline
$\alpha = \infty  $&$ 1 $&$ 1 $&$ 1 $&$ 1 $&$ 0 $&$ 0  $&$0$&$0 $&$ 0 $ \\
\hline
\end{tabular}
\end{center}
and we see that  $w(\Gamma)$ 
corresponds to minus the number of bound states of $H_\alpha$.

\subsection{The dimension $n=1$ with $\delta$-interaction}

The classification of self-adjoint extensions defining point interactions
is more complicated in one dimension. Also here one starts with the
Laplacian restricted to a subspace of functions which vanish at $0$
but there are more possibilities. We refer the reader to \cite{AGHH}
for the details considering in this section the family of extensions $H_\alpha$
called $\delta$-interactions. Here
the parameter $\alpha\in \R\cup\{\infty\}$ of the extension describes
the boundary condition $\psi'(0_+)-\psi'(0_-) = \alpha \psi(0)$ which can
be formally interpreted as arising from a potential
$V=\alpha\delta$ where $\delta$ is the Dirac $\delta$-function at $0$.
The extension for $\alpha=0$ is the free Laplace operator and the
extension $\alpha=\infty$ the Laplacian 
 (or rather the direct sum of two Laplacians) 
with Dirichlet boundary
conditions at $0$. The extensions $H_\alpha$ have a
single eigenvalue if $\alpha<0$ and do not have any eigenvalue if 
$\alpha \in [0,\infty]$ . Furthermore, the wave
operator $\Omega_-^\alpha$ for the couple $(H_\alpha, -\Delta)$ acts on $\psi
\in L^2(\R)$ as 
\begin{equation*}
\big[(\Omega_-^\alpha-1)\psi\big](x)=s-\lim_{R\to \infty}(2\pi)^{-1/2}
\int_{k\leq R}\de k \int_{\S^0} \de \omega
\;\!\frac{-i\alpha}{2k+i\alpha}e^{ik|x|} \;\!\hat{\psi}(k\omega)\ ,
\end{equation*}
where $\hat{\psi}$ denotes the $1$-dimensional Fourier transform of $\psi$. 
By rewriting this operator in terms of $-\Delta$ 
and $A$ one obtains:
\begin{equation*}
\Omega_-^\alpha-1 = \left[\frac{1}{2}\left(1+\tanh(\pi A) + i \big(\cosh(\pi A)\big)^{-1}\right)\right]
\left\{\frac{-2i \alpha}{2\sqrt{-\Delta}+i\alpha} \right\} P_0\ .
\end{equation*}
Thus we get the following results for the functions $\Gamma_i$ and
their contribution to the winding number. Let us set
\begin{equation*}
r(\xi) =  -\tanh(\pi \xi) - i\big(\cosh(\pi \xi)\big)^{-1}
\end{equation*}
and
\begin{equation*}
s^\alpha(\xi) = \frac{2\sqrt{\xi}-i\alpha}{2\sqrt{\xi}+i\alpha}\ .
\end{equation*}
Then
\begin{center}
\begin{tabular}{|c|c|c|c|c|c|c|c|c|c|}
\hline
 & $\Gamma_1$ & $\Gamma_2$ & $\Gamma_3$ & $\Gamma_4$  & $w_1$
& $w_2$ & $w_3$ & $w_4$ & $w(\Gamma)$ \\ \hline\hline 
$\alpha < 0$ &$ r $&$ s^\alpha $&$ 1 $&$ 1 $&$ -\frac{1}{2} $&$ -\frac{1}{2}  $&$
0 $&$0 $&$ -1 $ \\\hline
$\alpha = 0  $&$  1 $&$ 1 $&$ 1 $&$ 1 $&$ 0  $&$0$& $
0$&$0 $&$ 0 $ \\\hline
$\alpha > 0  $&$ r $&$ s^\alpha $&$ 1 $&$ 1 $&$ -\frac{1}{2} $&$ \frac{1}{2}  $&$
0 $&$0 $&$ 0 $ \\ \hline
$\alpha = \infty  $&$  r $&$ -1 $&$ r $&$ 1 $&$ -\frac{1}{2} $&$0$& $
\frac{1}{2} $&$0 $&$ 0 $ \\
\hline
\end{tabular}
\end{center}
and we see that  $w(\Gamma)$ 
corresponds to minus the number of bound states of $H_\alpha$.

\subsection{The dimension $n=1$ with $\delta'$-interaction}

Let us finally consider the family of extensions called $\delta'$-interaction. 
As in \cite{AGHH} we use $\beta\in\R\cup\{\infty\}$ 
for the parameter of the self-adjoint extension which describes
the boundary condition $\psi(0_+)-\psi(0_-) = \beta \psi'(0)$. This can
be formally interpreted as arising from a potential
$V=\beta\delta'$. The extension for $\beta=0$ is the free Laplace operator and the
extension $\beta=\infty$ the Laplacian (or rather the direct sum of
two Laplacians) with Neumann boundary
conditions at $0$. The operator $H_\beta$ possesses a single
eigenvalue if $\beta < 0$ of value $-4\beta^{-2}$ but no
eigenvalue if $\beta \in [0,\infty]$.
The wave operator $\Omega_-^\beta$ for the couple $(H_\beta, -\Delta)$ acts on any $\psi \in L^2(\R)$ as
\begin{equation*}
\big[(\Omega_-^\beta-1)\psi\big](x)=s-\lim_{R\to \infty}(2\pi)^{-1/2}
\int_{k\leq R}\de k \int_{\S^0} \de \omega \;\!\frac{i\beta k \omega}{2-i\beta k}\vartheta(x,k)
\;\!\hat{\psi}(k\omega)\ ,
\end{equation*}
where $\hat{\psi}$ denotes the $1$-dimensional Fourier transform of $\psi$ and with 
$\vartheta(x,k) = e^{ikx}$ for $x>0$ and $\vartheta(x,k)=-e^{-ikx}$ for $x<0$.

It is easily observed that the action of $\Omega_-^\beta-1$ on any symmetric ({\it i.e.}~even) function is trivial.
Moreover, this operator can be rewritten as 
\begin{equation*}
\Omega_-^\beta-1 = 
\left[\frac{1}{2}\left(1+\tanh(\pi A)-i\big(\cosh(\pi A)\big)^{-1}
\right)\right]
\left\{\frac{2i \beta \sqrt{-\Delta}}{2-i\beta\sqrt{-\Delta}} \right\} P_1\ .
\end{equation*}
We get the following results for the functions $\Gamma_i$ and
their contribution to the winding number. Let us set
\begin{equation*}
r(\xi) =  -\tanh(\pi \xi)
+i\big(\cosh(\pi \xi)\big)^{-1}
\end{equation*}
and
\begin{equation*}
s^\beta(\xi) =  
\frac{2 + i \beta \sqrt{\xi}}{2 - i \beta \sqrt{\xi}}\ .
\end{equation*}
Then
\begin{center}
\begin{tabular}{|c|c|c|c|c|c|c|c|c|c|}
\hline
 & $\Gamma_1$ & $\Gamma_2$ & $\Gamma_3$ & $\Gamma_4$  & $w_1$
& $w_2$ & $w_3$ & $w_4$ & $w(\Gamma)$ \\ \hline\hline 
$\beta < 0$ &$ 1 $&$ s^\beta $&$ r $&$ 1 $&$ 0 $&$ -\frac{1}{2}  $&$
-\frac{1}{2}$&$0 $&$ -1 $ \\\hline
$\beta = 0  $&$ 1 $&$ 1 $&$ 1 $&$ 1 $&$ 0 $&$ 0 $&$ 0 $&$0 $&$ 0 $ \\
\hline
$\beta > 0  $&$ 1 $&$ s^\beta $&$ r $&$ 1 $&$ 0 $&$ \frac{1}{2}  $&$
-\frac{1}{2}$&$0 $&$ 0 $ \\
\hline
$\beta = \infty  $&$  r $&$ -1 $&$ r $&$ 1 $&$ \frac{1}{2}  $&$0$& $
-\frac{1}{2}$&$0 $&$ 0 $ \\\hline
\end{tabular}
\end{center}
which verifies again the theorem.

\begin{Remark} 
{\rm 
It can be observed that the fonctions $\varphi$ and $\eta$ for the operators
$\Omega_-^\alpha-1$ are always given by $\hbox{$\frac{1}{2}$}(1-r)$ and $s^\alpha-1$
respectively. A straightforward calculation shows that the operators $\Omega_+-1$ can also 
be rewritten in the form $\varphi(A)\eta(-\Delta)P$ with $\varphi=\hbox{$\frac{1}{2}$}(1+r)$
and $\eta=\overline{s^\alpha -1}$.

The explicit formulae obtained for the wave
operator allow us to observe a symmetry among the models in
one and three dimensions. We see exactly the same
formulas for $\varphi$ and $\eta$ in the case of
the $\delta$-interaction in $n=3$ and the
$\delta'$-interaction in $n=1$, provided we set 
$2\pi\alpha=\beta^{-1}$. From the $C^*$-algebraic point of view the
fact that $\varphi$ and $\eta$ are the same 
means that the wave operators for the models 
are just two different
representations of the same element of an abstract $C^*$-algebra. }
\end{Remark}

\section*{Acknowledgements}

The second author thanks B. Helffer for a two weeks invitation to
Orsay where a substantial part of the present
calculations was performed. 
This stay was made possible thanks to the European Research Network
"Postdoctoral  
Training Program in Mathematical Analysis of Large Quantum Systems"
with contract number HPRN-CT-2002-00277.

\end{document}